\documentclass{aastex701}

\usepackage[utf8]{inputenc}
\usepackage[T1]{fontenc}
\usepackage{threeparttable}

\usepackage{amsmath}
\usepackage{microtype}

\begin{document}

\title{Unveiling the X-ray properties of the eclipsing Cataclysmic Variable UU~Aqr: spatially and spectrally-resolved two-component emission}

\author[0000-0002-2413-9301]{Nazma Islam}
\affil{Center for Space Science and Technology, University of Maryland, Baltimore County, 1000 Hilltop Circle, Baltimore, MD 21250, USA}
\affil{X-ray Astrophysics Laboratory, NASA Goddard Space Flight Center, Greenbelt, MD 20771, USA}
\email{nislam@umbc.edu}

\author[0000-0002-8286-8094]{Koji Mukai}
\affil{Center for Space Science and Technology, University of Maryland, Baltimore County, 1000 Hilltop Circle, Baltimore, MD 21250, USA}
\affil{CRESST II and X-ray Astrophysics Laboratory, NASA Goddard Space Flight Center, Greenbelt, MD 20771, USA}
\email{Koji.Mukai@umbc.edu}

\author[0000-0002-3331-7595]{Maurice A. Leutenegger}
\affil{X-ray Astrophysics Laboratory, NASA Goddard Space Flight Center, Greenbelt, MD 20771, USA}
\email{maurice.a.leutenegger@nasa.gov}

\author{Gabriel W. Pratt}
\affil{Université Paris-Saclay, Université Paris Cité, CEA, CNRS, AIM de Paris-Saclay, 91191 Gif-sur-Yvette, France}
\email{gabriel.pratt@cea.fr}

\correspondingauthor{Nazma Islam}
\email{nislam@umbc.edu}

\begin{abstract}

Non-magnetic Cataclysmic Variables (CVs) show two distinct X-ray components: a hard, optically thin component and a soft, optically thick, blackbody-like component, both produced in the boundary layer between the accretion disk and the White Dwarf (WD). An additional soft component originating from a more extended region has been reported in few CVs. In a short Chandra exposure, we identified a tentative X-ray eclipse in UU~Aqr, a non-magnetic CV which shows deep optical eclipses. Using observations with the Nuclear Spectroscopic Telescope Array (NuSTAR) and the XMM-Newton, we detect total eclipses in the orbital intensity profiles of this system in the hard X-ray band (3–10 keV with XMM and 3–25 keV with NuSTAR). However, the soft X-ray band (0.3–2.0 keV) shows no evidence of an eclipse. Detailed eclipse modeling, energy-resolved power spectral analysis and broadband spectral modeling indicate that the hard absorbed X-ray emission originates from a compact region near the WD, such as a boundary layer, while the soft, unabsorbed and un-eclipsed X-ray emission originates in an extended region. Neither scattering of hard X-rays nor colliding winds can account for the observed un-eclipsed soft emission. We instead propose that this component is produced by shocks within vertically extended, radiatively driven accretion-disk winds. We also provide new estimates on the emitting regions, mass and radius of the WD and the donor star using eclipse modeling. 

\end{abstract}

\keywords{}

\section{Introduction}

Non-magnetic CVs consist of a WD accreting matter from a late-type companion star. In the absence of a strong magnetic field, the accreting material forms a hot accretion disk around the WD, emitting predominantly in the infrared, optical, and UV bands \citep{patterson1985b}. 
In a subclass of non-magnetic CVs, called the Dwarf Novae (DNe), the mass transfer rate from the companion star to the disk is not high enough to maintain the accretion disk in a stable state. These systems spend the majority of the time in quiescence, during which the transferred mass accumulates in the disk, with only a fraction being accreted onto the WD. Thermal instability of the disk triggers occasional outbursts, during which it becomes hot and luminous, and the accretion rate onto the WD increases \citep{hameury2020,lasota2001}. The other subclass of non-magnetic CVs are called nova-like systems (also called nova-like variables) because they resemble systems that recently underwent classical nova eruptions\footnote{Classical nova eruptions are much more dramatic brightening of accreting WD systems, caused by thermonuclear runaways on the WD surface\citep{chomiuk2021, Islam2024}}. In nova-like systems, the mass transfer rate is high enough to keep the accretion disk in a steady, high-accretion, high-flux state analogous to disks in DNe in outbursts. Some systems occasionally exhibit low states, attributed to temporary reduction in the mass transfer rate to the accretion disk, rather than to an instability in the disk.
\par
The primary site of X-ray emission in non-magnetic CVs is considered to be the boundary layer between the Keplerian part of the accretion disk and the slowly rotating WD. This boundary layer is optically thin in low states, producing predominantly hard X-rays ($>$0.5 keV)  \citep{patterson1985a}, and becomes optically thick in high states and emits mainly soft X-rays ($<$0.5 keV)  \citep{patterson1985b}. Observations of the DN SS Cyg during outburst reveal the sudden emergence of a bright EUV/soft X-ray component, indicating a transition to an optically thick boundary layer \citep{wheatley2003a}. This transition was accompanied by a sharp but incomplete suppression of the hard X-ray flux. \citet{patterson1985b} suggested that the residual hard X-rays originate from an optically thin ``skin" of the boundary layer, whereas \citet{ishida2009} proposed an alternative origin involving a hot corona above the optically thick disk photosphere. The origin of hard X-ray emission in DNe in outbursts still remains a mystery.
\par
An important question is whether the boundary layer in nova-like systems is analogous to that in DNe during outburst. While DNe are typically more luminous in hard X-rays during (optically defined) quiescence than during outburst \citep{fertig2011}, no such pattern is evident in nova-like systems \citep{zemko2014}. It is likely that the accretion disks (and the boundary layers) of DNe in outburst never settle into a steady state, in contrast to the steady state configuration of accretion disks and boundary layers in nova-like systems, which might explain their different X-ray emission characteristics. 
\par
A puzzling soft, un-eclipsed X-ray component is observed in a few high-inclination DNe in outburst and in nova-like systems. In these systems, where optical eclipses occur from the occultation of the WD by the secondary star, analogous X-ray eclipses are expected from the occultation of the boundary layer emission. This is confirmed for the X-ray emission from quiescent eclipsing DNe, as shown by observations of OY Car \citep{pra99a, wheatley2003b}, HT Cas \citep{wood95,nucita2009}, and Z Cha \citep{vant97,nucita2011}. However in OY Car during outburst, the soft X-ray component remains un-eclipsed, indicating that the soft X-ray emission originates from an extended region \citep{naylor1988,pra99b}. The XMM observation of the nova-like system UX~UMa revealed both an eclipsed, absorbed hard X-ray component and an un-eclipsed, unabsorbed soft X-ray component \citep{pratt2004}. UX~UMa is the only nova-like system which exhibits an eclipsed hard X-ray component together with an un-eclipsed soft X-ray component, until the present work.
\par
UU Aquarii (UU~Aqr) is a optically bright nova-like system with $\langle V\rangle \sim$ 13.5 and an orbital period of 3.94 hr. The system exhibits eclipses in optical as well as displays long-term optical intensity variations on a timescale of about four years. An orbital inclination of $78^\circ$ has been estimated from optical eclipses \citep{baptista1994}. The distance to the source reported in Gaia EDR3 is 257.3$\pm$1.3 pc \citep{bailerjones2021}.
\par
In this manuscript, we present results from Chandra, NuSTAR, and XMM observations of UU~Aqr. The structure of the manuscript is as follows: Section 2 describes the observational data and analysis methods, including the construction of energy-resolved orbital intensity profiles and power spectra, and broadband X-ray spectral modeling. In Section 3, we discuss the results of the energy resolved orbital intensity profiles and their implications for understanding the origin of the UV, soft and hard X-ray emission components. We further use the eclipse modeling to constrain the size of the X-ray and UV emitting regions as well as provide estimates on the mass and radius of the secondary star and WD. We propose a physical interpretation for the soft and hard X-ray emission in UU~Aqr and discuss its wider significance for understanding the origin of X-ray emission from non-magnetic CVs.

\section{X-ray and UV Observations and Analysis}

\begin{figure}
    \centering
    \includegraphics[scale=0.5]{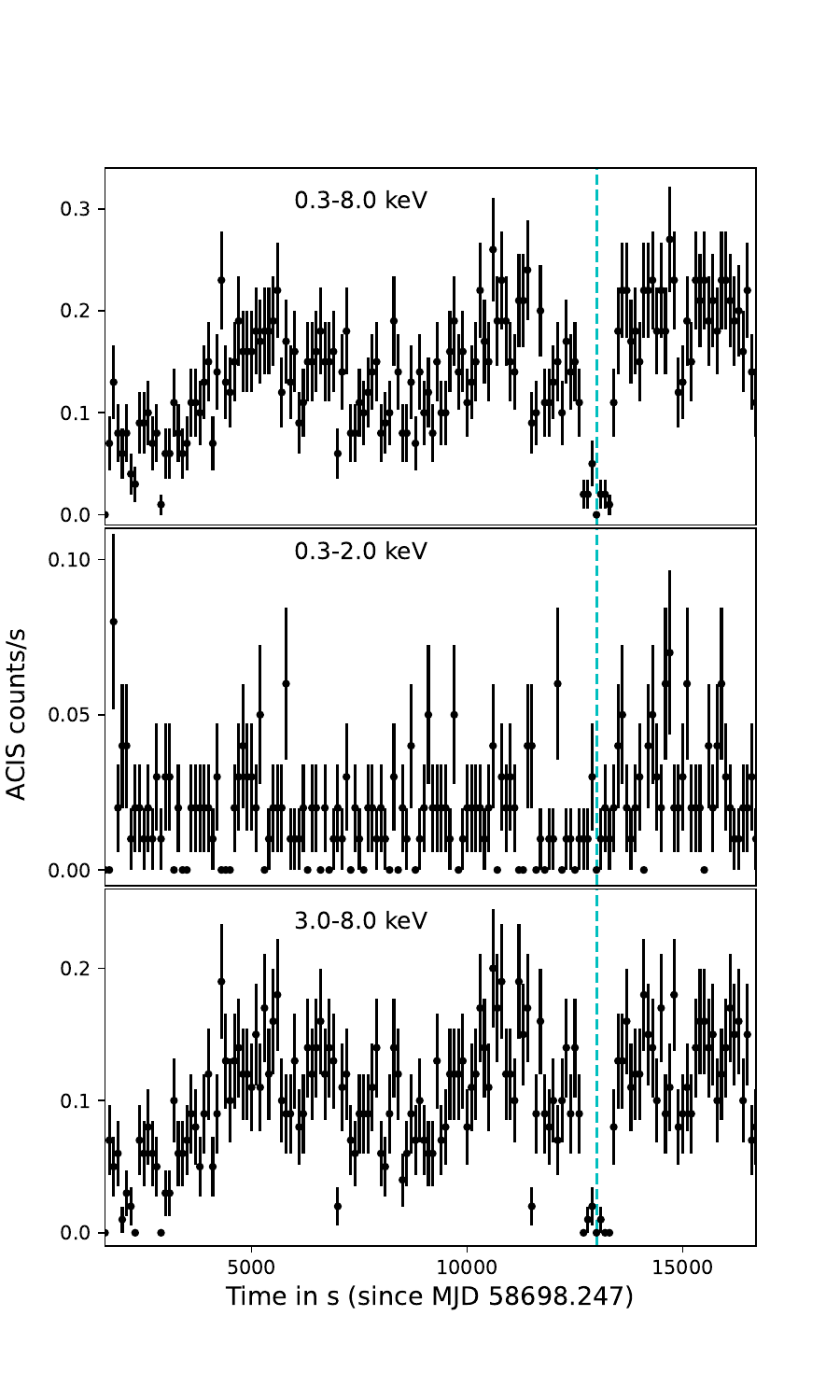}
    \includegraphics[scale=0.5]{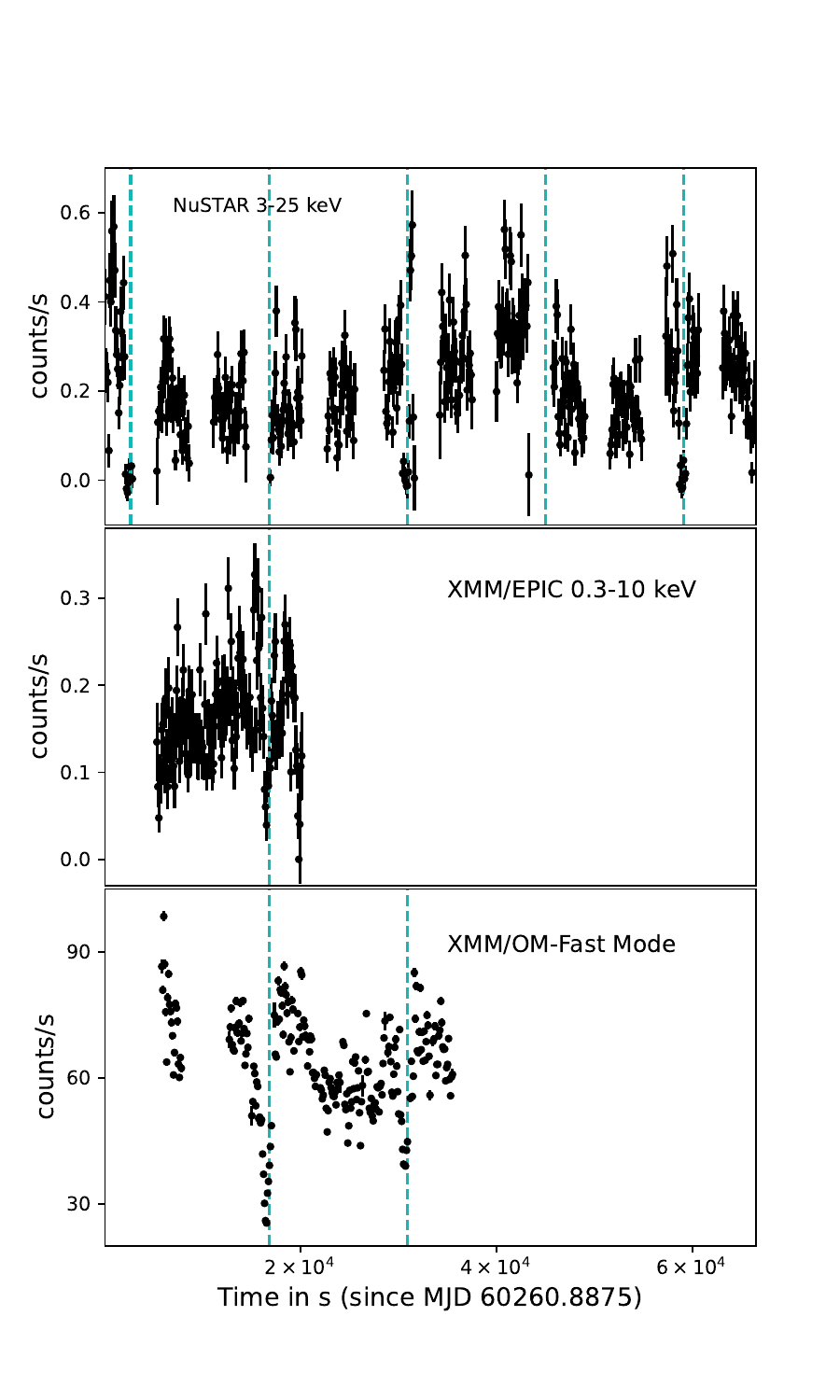}
    \caption{{\it Left panel}: Chandra-ACIS background subtracted lightcurve of UU~Aqr, covering one orbital cycle, in the 0.3--8 keV (top panel), 0.3-2.0 keV (middle panel) and 3.0-8.0 keV (bottom panel) energy bands. 
    {\it Right panel}: NuSTAR FPMA+FPMB background subtracted lightcurve of UU~Aqr in 3--25 keV (top panel) and XMM EPIC (MOS1 + MOS2 + PN) background subtracted lightcurve in 0.3--10 keV (middle panel) and XMM OM-Fast Mode using UVW1 filter (bottom panel), binned with 100 s. The NuSTAR observation covered about five orbital cycles of the source, the XMM observation covered one orbital cycle for EPIC and 2 orbital cycles for OM-Fast Mode. The cyan dashed lines mark the expected optical eclipses using the ephemeris given in \cite{baptista1994} and an orbital period of 0.163580429 d.}
    \label{lc}
\end{figure}

UU~Aqr was observed with Chandra-ACIS, as part of an observational campaign to study X-ray properties of nova-like variables (Islam N. et al. in prep) on August 3, 2019 for 14 ks in Timed Exposure Mode using Faint Telemetry format (ObsID: 21280). The target was imaged on the ACIS-S3 chip. We reduced the data using Chandra Interactive Analysis of Observations (CIAO) software ver. 4.17 (CALDBver. 4.12.2). For carrying out the timing analysis, we transformed the event times of arrival to the solar system barycenter using the {\tt barycen} tool. The left panel of Figure \ref{lc} shows the background subtracted lightcurves in energy bands 0.3-8.0 keV, 0.3-2.0 keV and 3.0-8.0 keV. We detect a low count-rate episode in both the 0.3–8.0 keV and 3.0–8.0 keV energy bands, likely an X-ray eclipse (indicated by the dashed lines), at the epoch consistent with the ephemeris of the optical eclipses reported by \cite{baptista1994}. However, due to limited photon statistics and low sensitivity of Chandra-ACIS below 1 keV due to contaminant build-up \citep{marshall2004, dell2017}, it is unclear whether the X-ray eclipse is partial or complete. No X-ray eclipse is evident in the 0.3–2.0 keV lightcurve. However the low photon counts make this non-detection inconclusive.
\par
Motivated by the indication of an X-ray eclipse in the Chandra observation, we carried out joint NuSTAR and XMM-Newton observation of UU~Aqr (P.I: N.Islam). NuSTAR is a hard X-ray telescope operating in 3--79 keV energy band \citep{harrison2013}. It carries two co-aligned grazing incidence Wolter I imaging telescopes that focus onto two independent Focal Plane Modules, FPMA and FPMB. The NuSTAR observation of UU~Aqr was carried out on 2023-02-13 with an effective exposure of 93 ks. The NuSTAR data was reduced and analyzed using NuSTAR Data Analysis Software (NuSTARDAS) v.2.1.4 package provided under HEAsoft v.6.34 and calibration files 2024-10-01. The event files were reprocessed using {\tt nupipeline} using the standard filtering procedure and the default screening criteria. The event times were corrected to the solar system barycenter using nuproducts and the FTOOL {\tt barycorr} with the DE200 solar system ephemeris. The source spectra, response matrices, ancillary response files and energy-resolved lightcurves were extracted in SCIENCE mode (01) from a circular region of radius 60\arcsec\ centered on the source using {\tt nuproducts}. The background spectra and lightcurves were extracted from a similar circular region in a source-free region on the same chip. The lightcurves were extracted in the energy-bands 3--25 keV, 3--10 keV and 10--25 keV. We did not use the energy band above 25 keV, beyond which the background count-rates dominate the source. 
\par
An XMM-Newton observation \citep{jansen2001} was carried out simultaneously with the NuSTAR observation, with an exposure of 22 ks (ObsID: 0930800101). The European Photon Imaging Camera (EPIC) PN and MOS cameras were operated in Full Window Imaging mode. The data were analyzed using SAS v21.0.0. We identified intervals of flaring particle background by extracting the single event (PATTERN == 0) 10-12 keV lightcurve for PN and MOS. The threshold for low steady background was determined to be for a RATE $\leq$ 0.4 counts/s for PN and RATE $\leq$ 0.5 counts/s for MOS. This removed the part of the observation after t = 22 ks as shown in the right middle panel of Figure \ref{lc}. The lightcurves and spectral files were extracted using the XMM SAS data analysis threads\footnote{https://www.cosmos.esa.int/web/xmm-newton/sas-threads}, which filtered out the background flares and applied the latest calibration files. A circular source region was selected with a radius of 60” centered on the source and a similar size background region was selected in a source-free region on the chip. We included an empirical correction to the EPIC effective area based on NuSTAR observations by applying the parameter {\tt applyabsfluxcor=yes} to the XMM SAS tool {\tt arfgen} as suggested by the XMM Science Operations Team \footnote{https://xmmweb.esac.esa.int/docs/documents/CAL-TN-0230-1-3.pdf}. Background subtracted lightcurves were extracted in the 0.3--10 keV, 0.3--2.0 keV and 3--10 keV energy bands. The Optical Monitor (OM) was operated in the Imaging and Fast Mode with UVW1 filter. The NuSTAR, XMM-EPIC and Chandra-ACIS source spectra were grouped to give a minimum of 30 counts in each spectral bin. 
\par
The right panel of Figure \ref{lc} presents the simultaneous background subtracted NuSTAR and XMM-EPIC lightcurves in the 3-25 keV and 0.3-10 keV bands respectively, along with the XMM OM-Fast UVW1 lightcurves. The NuSTAR light curves span roughly five orbital cycles, the XMM-EPIC lightcurves cover a single orbital cycle, and the XMM-OM-Fast lightcurves extend across two orbital cycles, with the expected eclipses from the optical ephemeris by \cite{baptista1994} indicated by cyan dashed lines. Outside the expected eclipse, some variability is present, but it is neither pronounced nor systematic across the intervals observed with XMM-EPIC and NuSTAR. We therefore assume that the spectral shape remains largely unchanged throughout the full NuSTAR observation, and that the cross-normalization constant adequately accounts for the minor variability between the XMM-EPIC and NuSTAR spectra.

\subsection{Orbital Intensity profiles}

\begin{figure}
    \centering
    \includegraphics[scale=0.7]{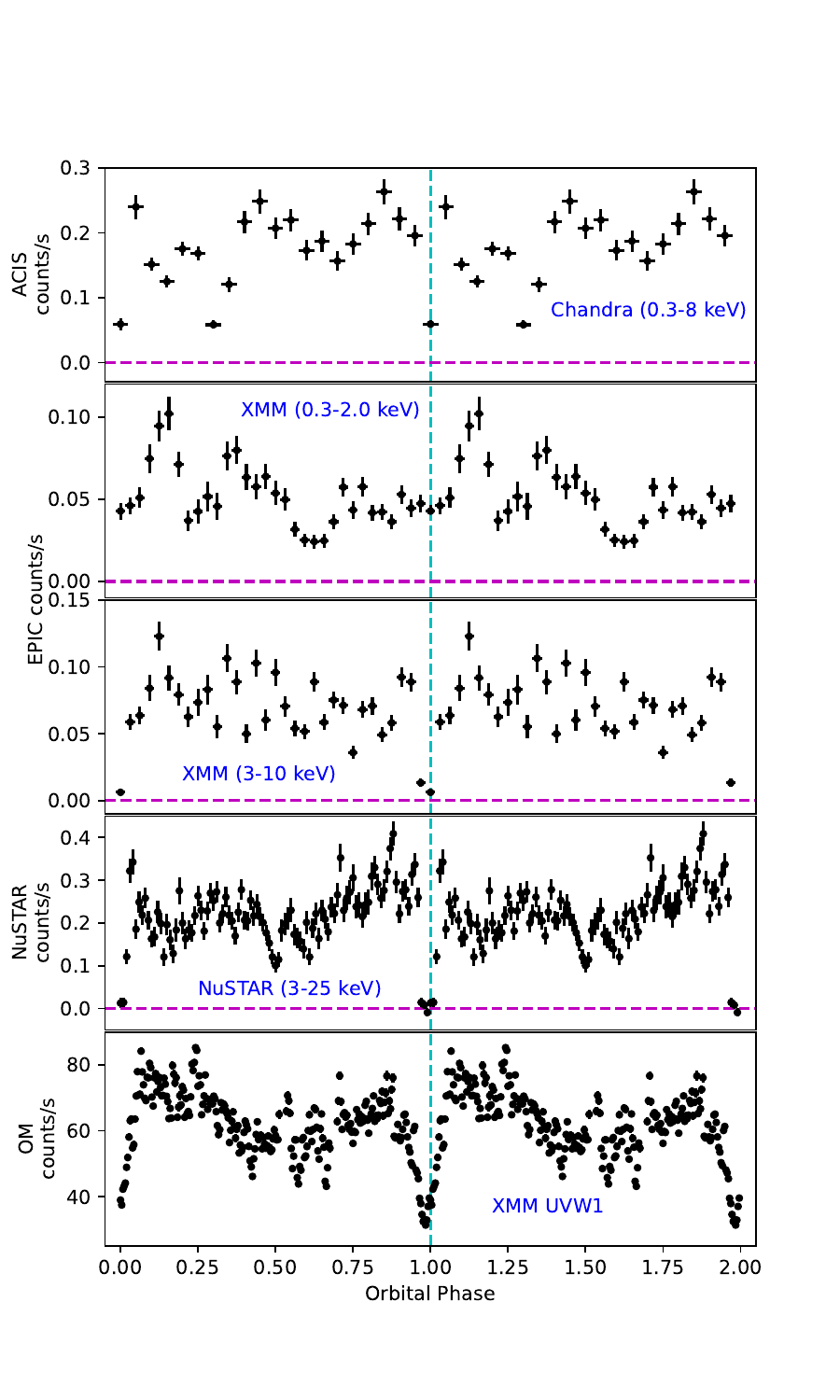}
    \caption{Orbital intensity profile of UU~Aqr constructed using background subtracted Chandra-ACIS lightcurves in 0.3--8.0 keV using 20 phasebins (top panel), energy-resolved background subtracted lightcurves of the XMM observation in 0.3--2.0 keV (second panel) and 3--10 keV using 32 phasebins (third panel), background subtracted NuSTAR lightcurves in 3--25 keV in 100 phasebins (fourth panel) and XMM OM-Fast UVW1 lightcurves in 100 phasebins (bottom panel). The orbital ephemeris used for folding the lightcurves is taken from the optical eclipses in \cite{baptista1994}, which mentions the mid-eclipse time T$_{0}$ = JD 2446,347.26657 and orbital period P = 0.163580429 d. The vertical cyan dashed line marks the mid-eclipse center and the horizontal magenta dashed line marks the zero count-rate of the X-ray observations.}
    \label{orb_phase}
\end{figure}

We constructed orbital intensity profiles by folding the 0.3--8 keV Chandra/ACIS lightcurve in 20 phasebins, the energy-resolved 0.3--2.0 keV and 3--10 keV XMM lightcurves in 32 phasebins, the 3--25 keV NuSTAR lightcurve in 100 phasebins and the XMM OM using UVW1 filter lightcurve in 300 phasebins. Figure \ref{orb_phase} displays these orbital intensity profiles folded using the ephemeris from \cite{baptista1994}, which was estimated from optical eclipses: mid-eclipse center T$_{0}$ = JD 2446,347.26657 and orbital period P = 0.163580429 d. The vertical cyan dashed line marks the mid-eclipse time of the optical eclipse using the above ephemeris.
\par
The orbital intensity profile constructed using the 0.3-8.0 keV Chandra-ACIS lightcurve, in the top panel of Figure \ref{orb_phase}, shows that phase zero (or phase of 1.0 for a repeated orbital cycle) corresponds to a low count-rate of 0.059$\pm$0.009 counts/sec, which likely indicates a partial X-ray eclipse. As mentioned in Section 2, the energy resolved Chandra-ACIS lightcurves (seen in the left panel of Figure \ref{lc}) suffer from low photon count-rates and the orbital intensity profiles constructed from them are therefore inconclusive in determining the nature of the X-ray eclipse (partial or total) or studying its energy dependence. 
\par
From the second panel of Figure \ref{orb_phase}, the XMM-EPIC soft X-ray band orbital intensity profile in 0.3--2.0 keV do not show the presence of an X-ray eclipse and has a count-rate of 0.043$\pm$0.005 counts/sec at the orbital phase of zero corresponding to the optical eclipse. However, we see a low count-rate of 0.006$\pm$0.002 corresponding to the orbital phase zero in the XMM hard X-ray band orbital intensity profile in 3--10 keV. This suggests that the X-ray emission from UU~Aqr consists of two components: a hard X-ray component which is eclipsed and is therefore located close to the WD, and a soft un-eclipsed X-ray component originating away from the orbital plane of the binary. This is similar to the X-ray behaviour of the eclipse observed in another nova-like system, UX~UMa \citep{pratt2004}. 
\par 
We notice that at orbital phase zero (which corresponds to the mid-center of the optical eclipse) of the NuSTAR 3--25 keV orbital intensity profile, the count-rate is zero, which indicates a complete hard X-ray eclipse in the system. In optical eclipse, the WD is expected to be exactly behind the secondary star. Since the optical and hard X-ray eclipse occur at the same orbital phase, the hard X-ray emitting region must be located close to the WD. 
\par
The eclipse observed in the UV band with the XMM OM-Fast Mode and UVW1 filter occurs at the same orbital phase as the optical and X-ray eclipses, but it is partial V-shaped rather than total flat bottomed eclipse seen in the hard X-rays. This suggests that the UV emission originates from an extended region in the orbital plane of the binary system, likely associated with the accretion disk. We present a detailed modeling of the eclipses in the next subsection.

\subsection{Fits to the Eclipse profile}

\begin{figure}
    \centering
    \includegraphics[scale=0.5]{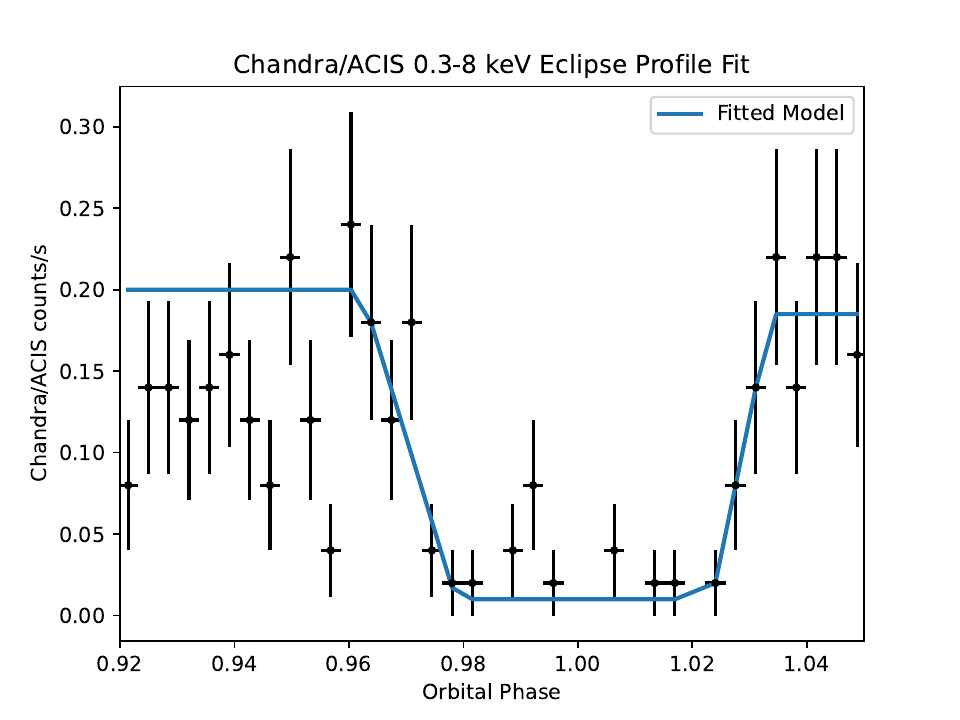}
    \includegraphics[scale=0.5]{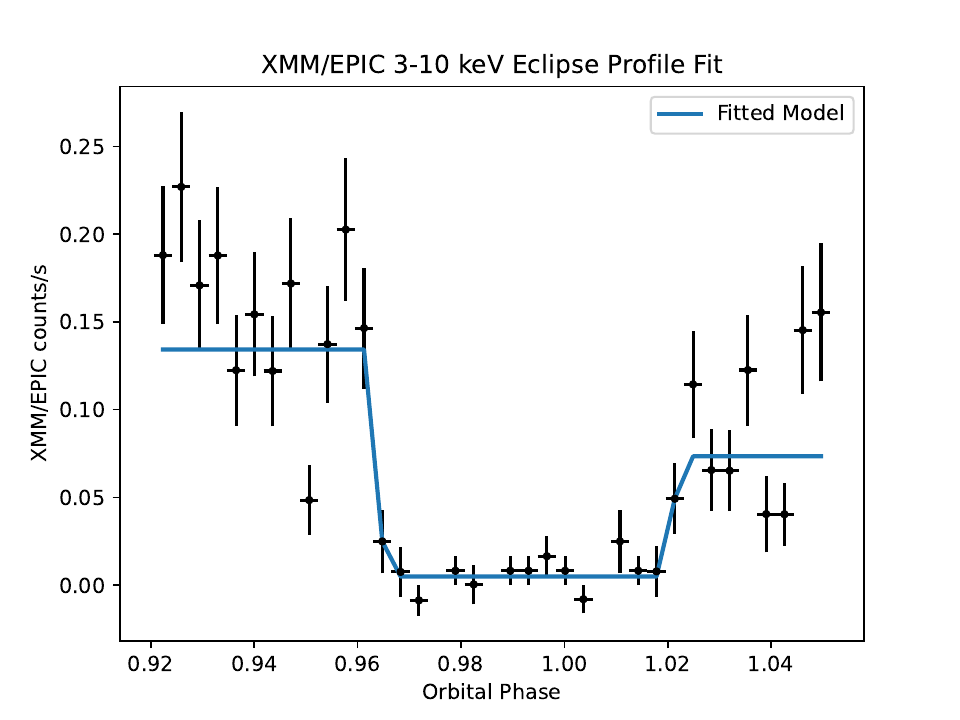}
    \includegraphics[scale=0.5]{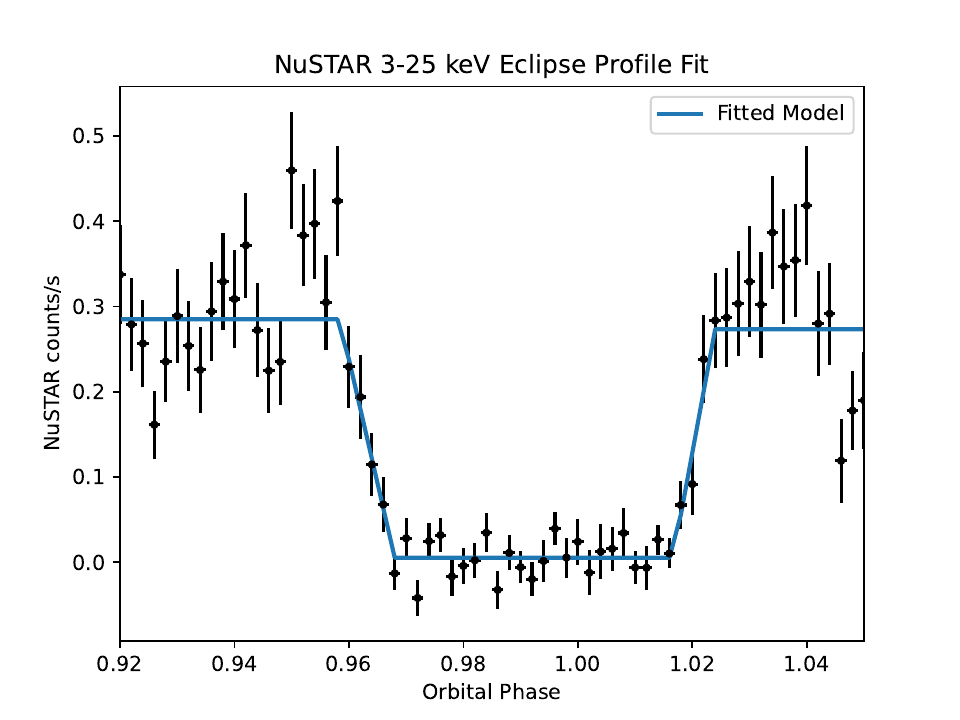}
    \includegraphics[scale=0.5]{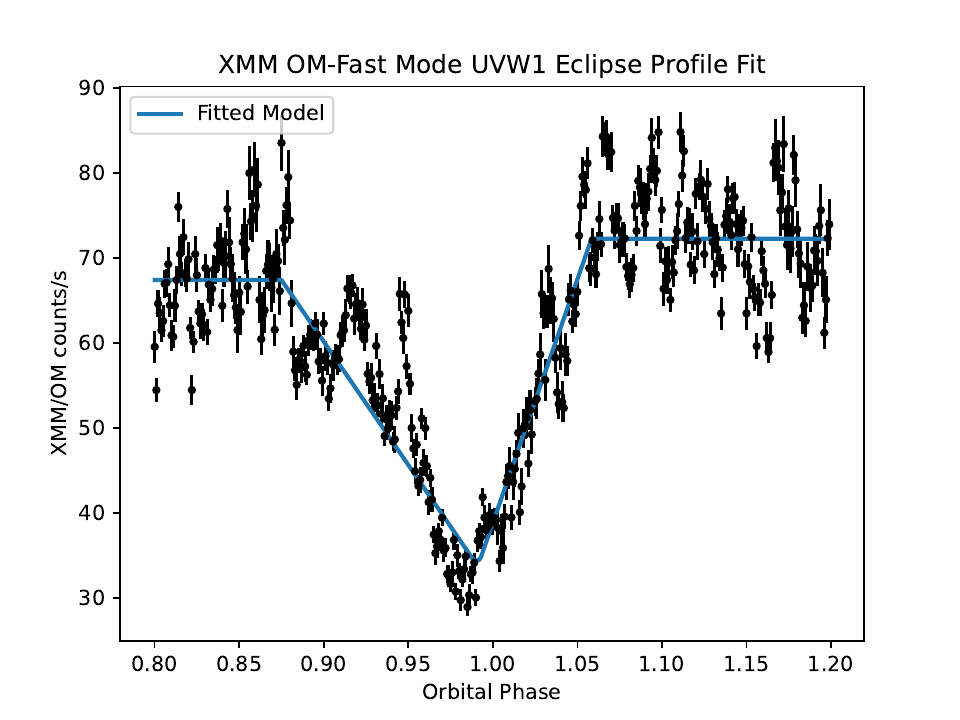}
    \caption{Orbital intensity profiles near the eclipse using Chandra 0.3--8 keV, XMM-EPIC 3--10 keV, NuSTAR 3--25 keV, and XMM-OM UVW1 lightcurves. These eclipse profiles are fitted with an asymmetric step and ramp function defined in Equation \ref{eclipse_eqn}. The results of the eclipse fits are reported in Table 1.}
    \label{eclipse_fit_asym}
\end{figure}

The eclipses observed in the orbital intensity profiles of Chandra, hard X-rays with XMM-EPIC and NuSTAR, and in XMM-OM with UVW1 filters in Figure \ref{orb_phase} are modeled using an asymmetric step and ramp function with the functional form shown in Equation \ref{eclipse_eqn} and fit with the Trust Region Reflective (trf) algorithm in {\tt scipy} (which is a bounded nonlinear least-squares minimization algorithm). The count-rates are assumed to remain constant before ingress, during eclipse and after egress, with a linear variation during the ingress and egress transition. The parameters of the model are as follows: the count-rate before ingress is $C_{\rm{out,ing}}$, the orbital phases corresponding to the start and end of ingress are $\phi_{\rm{in,start}}$ and $\phi_{\rm{in,end}}$ respectively, the count-rate inside the eclipse is $C_{\rm{in}}$, the orbital phases corresponding to the start and end of egress are $\phi_{\rm{eg,start}}$ and $\phi_{\rm{eg,end}}$ respectively and the count-rate after the egress is $C_{\rm{out,eg}}$. 

 \begin{equation}
    C(\phi)=
        \begin{cases}
            C_{out,ing} & \phi < \phi_{in,start} \\
            C_{out,ing} - (\frac{(C_{out,ing} - C_{in})(\phi - \phi_{in,start})}{\phi_{in,end} - \phi_{in,start} }) & \phi_{in,start} < \phi < \phi_{in,end} \\
            C_{in} & \phi_{in,end} < \phi < \phi_{eg,start} \\
            C_{in} + (\frac{(C_{out,eg} - C_{in})(\phi - \phi_{eg,start})}{\phi_{eg,end} - \phi_{eg,start} }) & \phi_{eg,start} < \phi < \phi_{eg,end} \\
            C_{out,eg} & \phi < \phi_{eg,end} \\
        \end{cases}
\label{eclipse_eqn}        
\end{equation}

 To carry out eclipse fitting with the above model, we used 500 phasebins for the NuSTAR lightcurves and fit the eclipse profile from orbital phase 0.92 to 1.05. We note that the orbital intensity profiles from the NuSTAR lightcurves are constructed by folding and averaging over 5 orbital cycles. Since the Chandra and the XMM lightcurves in 3--10 keV cover only one orbital cycle, we converted the 10 s binned lightcurves into an orbital intensity profile and fit them in the orbital phase 0.92 to 1.05. The photon limited statistics of the Chandra observations yield larger error-bars on the orbital intenstiy profiles, leading to correspondingly large uncertainties on the parameters of the eclipse fits. The eclipse profile for both the NuSTAR and XMM-EPIC 3-10 keV is characterized by a flat-bottomed shape and distinct, sharp ingress and egress transitions. The XMM-OM UVW1 orbital intensity profile is binned by 1000 phasebins and the eclipse fits are carried out in orbital phase 0.8 to 1.2. The eclipse profile in the XMM-OM UVW1 filter is V-shaped, indicating a shallow and partial UV eclipse, with gradual ingress and egress. Table 1 provides the results of fits of the eclipse profiles along with their errors calculated at 1$\sigma$ confidence level, and Figure \ref{eclipse_fit_asym} displays the best fits to the eclipse profile using Equation \ref{eclipse_eqn}. 

\begin{table*}
\centering
\begin{threeparttable}
\caption{Results of the fits to the eclipse profile using an asymmetric step and ramp function defined by Equation \ref{eclipse_eqn}, along with their errors calculated at 1$\sigma$ confidence level.}
\begin{tabular}{l c c c c c}
\hline
Parameter & NuSTAR & XMM 3-10 keV & Chandra & XMM-OM \\
\hline
Count-rate outside eclipse ingress$^{\star}$ ($C_{\rm out,ing}$) & 0.29$\pm$0.02 & 0.13$\pm$0.01 & 0.20$\pm$0.02 & 67.3$\pm$0.7 \\
Count-rate outside eclipse egress$^{\star}$ ($C_{\rm out,eg}$) & 0.27$\pm$0.02 & 0.07$\pm$0.01 & 0.19$\pm$0.05 & 72.2$\pm$0.5 \\
Count-rate inside eclipse$^{\star}$ ($C_{\rm in}$) & 0.005$\pm$0.005 & 0.005$\pm$0.004 & 0.01$\pm$0.01 & 35$\pm$2 \\
Eclipse Ingress Start$^{\dagger}$ ($\phi_{in,start}$) & 0.958$\pm$0.002 & 0.9615$\pm$0.0002 & 0.962$\pm$0.007 & 0.875$\pm$0.004  \\
Eclipse Ingress End$^{\dagger}$ ($\phi_{in,end}$) & 0.968$\pm$0.001 & 0.9654$\pm$0.0009 & 0.979$\pm$0.003 & 0.989$\pm$0.007 \\
Derived Ingress Duration$^{\dagger}$ & 0.010$\pm$0.003 & 0.004$\pm$0.001 & 0.02$\pm$0.01 & 0.11$\pm$0.01 \\
Eclipse Egress Start$^{\dagger}$ ($\phi_{eg,start}$) & 1.017$\pm$0.001 & 1.018$\pm$0.002 & 1.023$\pm$0.002 & 0.992$\pm$0.004 \\
Eclipse Egress End$^{\dagger}$ ($\phi_{eg,end}$) & 1.024$\pm$0.002 & 1.023$\pm$0.003 & 1.034$\pm$0.007 & 1.058$\pm$0.003 \\
Derived Egress Duration$^{\dagger}$ & 0.007$\pm$0.003 & 0.005$\pm$0.005 & 0.011$\pm$0.009 & 0.066$\pm$0.007 \\
\hline
\end{tabular}
\begin{tablenotes}[flushleft]
\item $\star$ Units of counts/s.
\item $\dagger$ Units of orbital phase.
\end{tablenotes}
\end{threeparttable}
\end{table*}

\subsection{Power Spectrum}

\begin{figure}
    \centering
    \includegraphics[scale=0.5]{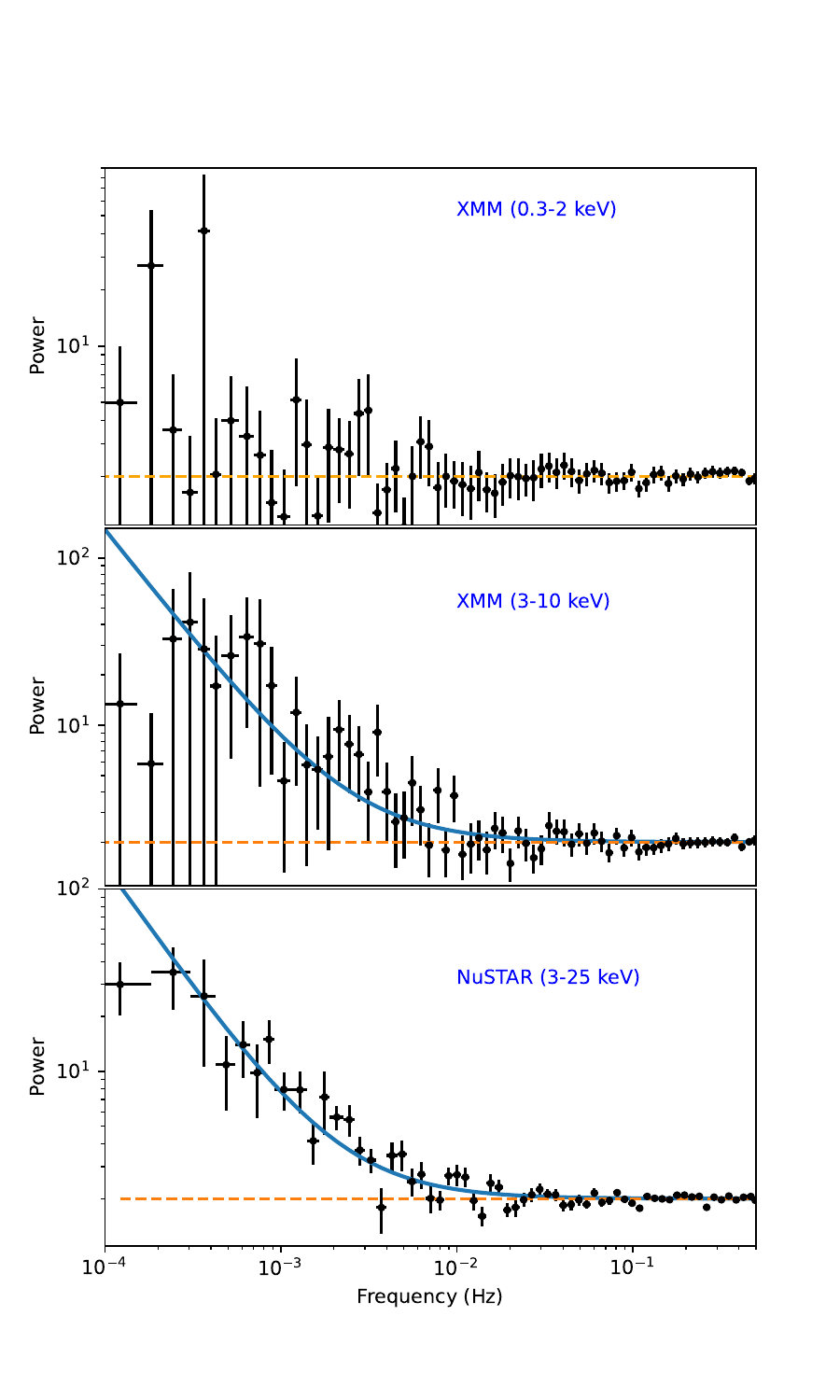}
    \includegraphics[scale=0.5]{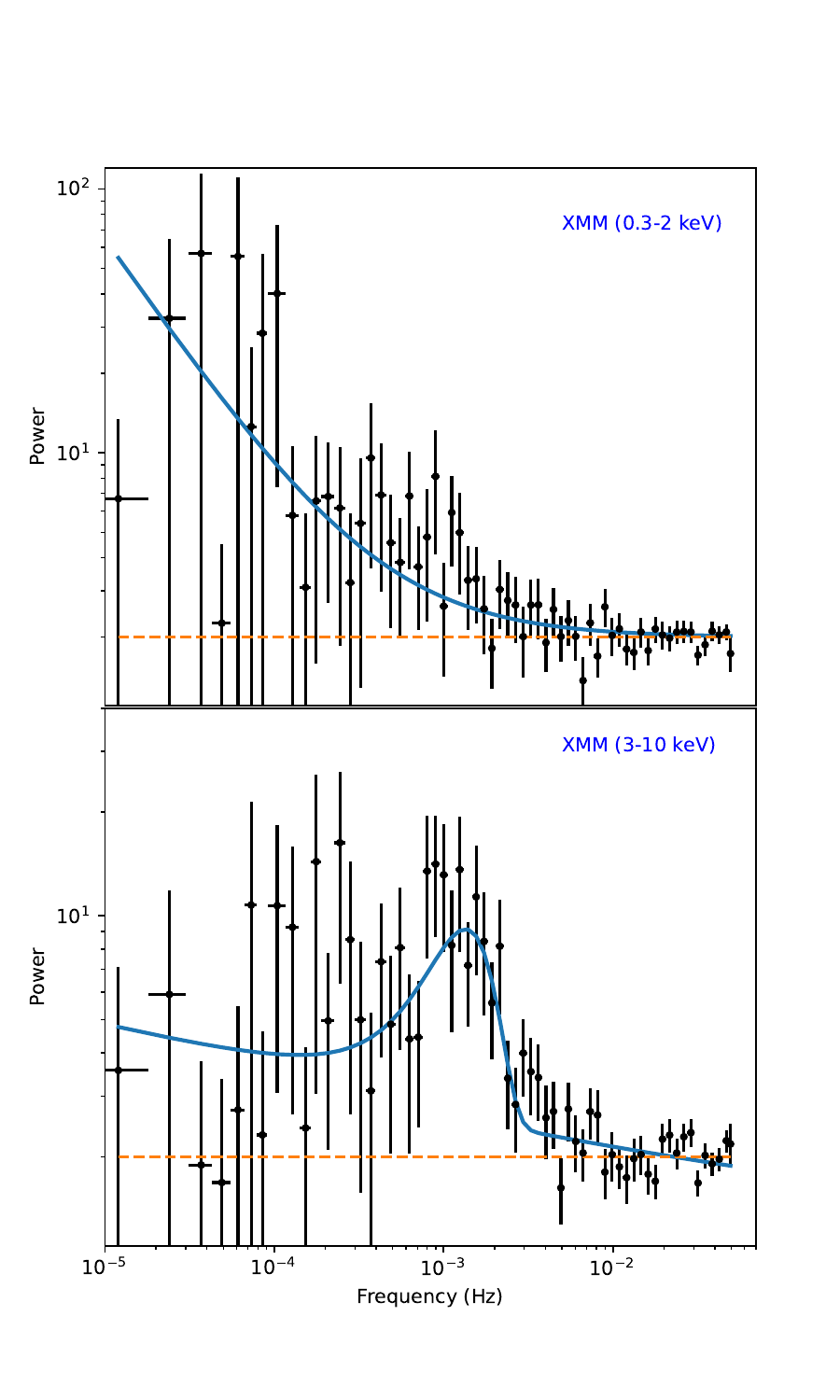}
    \caption{Power spectrum of UU~Aqr (left panel) and UX~UMa (right panel) constructed using energy resolved lightcurves. The power spectra constructed from the 3--10 keV XMM and 3--25 keV NuSTAR lightcurves for UU~Aqr is fit with a power-law model described in Section 2.3. The power spectrum constructed using 0.3--2 keV XMM lightcurves of UX~UMa is fit with a power-law model and the power spectrum from 3--10 keV XMM lightcurve is fit with a power-law + Gaussian model. The orange dashed lines denote the Poissonian power of 2.}
    \label{pow_spec}
\end{figure}

The orbital intensity profiles in Figure \ref{orb_phase} show the presence of two components to the X-ray emission: a hard, eclipsed component and a soft, un-eclipsed component, similar to another nova-like system UX~UMa \citep{pratt2004}. For UX~UMa, the hard, eclipsed X-ray component was thought to originate close to the WD, likely from the boundary layer between the WD and the accretion disk. The soft, un-eclipsed component was thought to originate away from the WD, and \citet{pratt2004} suggested scattering in the accretion disk winds as a potential source for this emission. 
\par
To further investigate the nature of this soft component in UU~Aqr, we constructed the power spectra of the energy resolved XMM lightcurves in 0.3-2 keV, 3-10 keV and NuSTAR lightcurves in 3-25 keV in the frequency range $10^{-4}$ to 0.5 Hz to evaluate their time variation. These power spectra were constructed using FTOOL {\tt powspec} with a bin time of 1 s and a geometric rebin factor to produce equi-spaced bins in log space. We used the Leahy normalization of the power spectra which ensures that white noise (Poisson noise) has a power level of 2.0 \citep{leahy1983}. The top left panel of Figure \ref{pow_spec} shows the power spectrum constructed using the XMM lightcurves in the soft energy band of 0.3--2 keV, and displays no significant time variation. The power spectra constructed from the XMM and NuSTAR lightcurves in hard energy bands of 3-10 keV and 3-25 keV respectively, are shown in the middle and bottom left panels of Figure \ref{pow_spec}. The power spectra derived from the hard-energy light curves exhibit a time variation that is described by a power-law model of the form 2 + b*$\mu ^{-a}$. Here, the constant term of 2 represents the expected white-noise level in the normalized power spectrum. The power-law indices for XMM 3--10 keV and NuSTAR 3-25 keV power spectra are 1.3$\pm$0.2 and 1.4$\pm$0.1 respectively. The power spectrum of the hard component shows a coherent variability within 100-1000 s whereas the power spectrum of the soft component shows stochastic variability and is consistent with the Poisson noise.
The characteristics of the power spectra constructed from the energy resolved lightcurves is similar to that seen in the power spectra of symbiotic binary CH Cyg, which consists of a giant star and a putative WD, using ASCA observations \citep{ezuka1998}. 
\par
We compare the time variation of the power spectra of the XMM-EPIC energy resolved lightcurves of UU~Aqr with that of the XMM observation of another nova-like system UX~UMa, which has been analyzed in \cite{pratt2004}. The power spectra of the soft and hard component of the XMM-EPIC lightcurves were not presented in \cite{pratt2004} and therefore we re-extracted the energy resolved lightcurves of the XMM-EPIC observation of UX~UMa (ObsID: 0084190201) in the soft band of 0.3-2 keV and hard band of 3-10 keV, using the same analysis procedure mentioned in \cite{pratt2004}. The right panel of Figure \ref{pow_spec} displays the power spectra of the energy resolved XMM-EPIC lightcurves of UX~UMa. This can be modeled by a power-law (the functional form described above) with an index of 0.9$\pm$0.1 in the soft X-ray band, and a Gaussian + power-law model with Gaussian center at 1.4$\pm$0.1 mHz, $\sigma$ of the Gaussian 0.6$\pm$0.1 mHz and power-law index of 0.2$\pm$0.1 in the hard X-ray band. We find that the power spectrum of both the soft and hard X-ray component exhibits coherent variability in the timescale of 100-1000 s, unlike that observed in the power spectra of UU~Aqr. 

\subsection{X-ray spectral Fitting}
\begin{figure}
    \centering
    \includegraphics[scale=0.3]{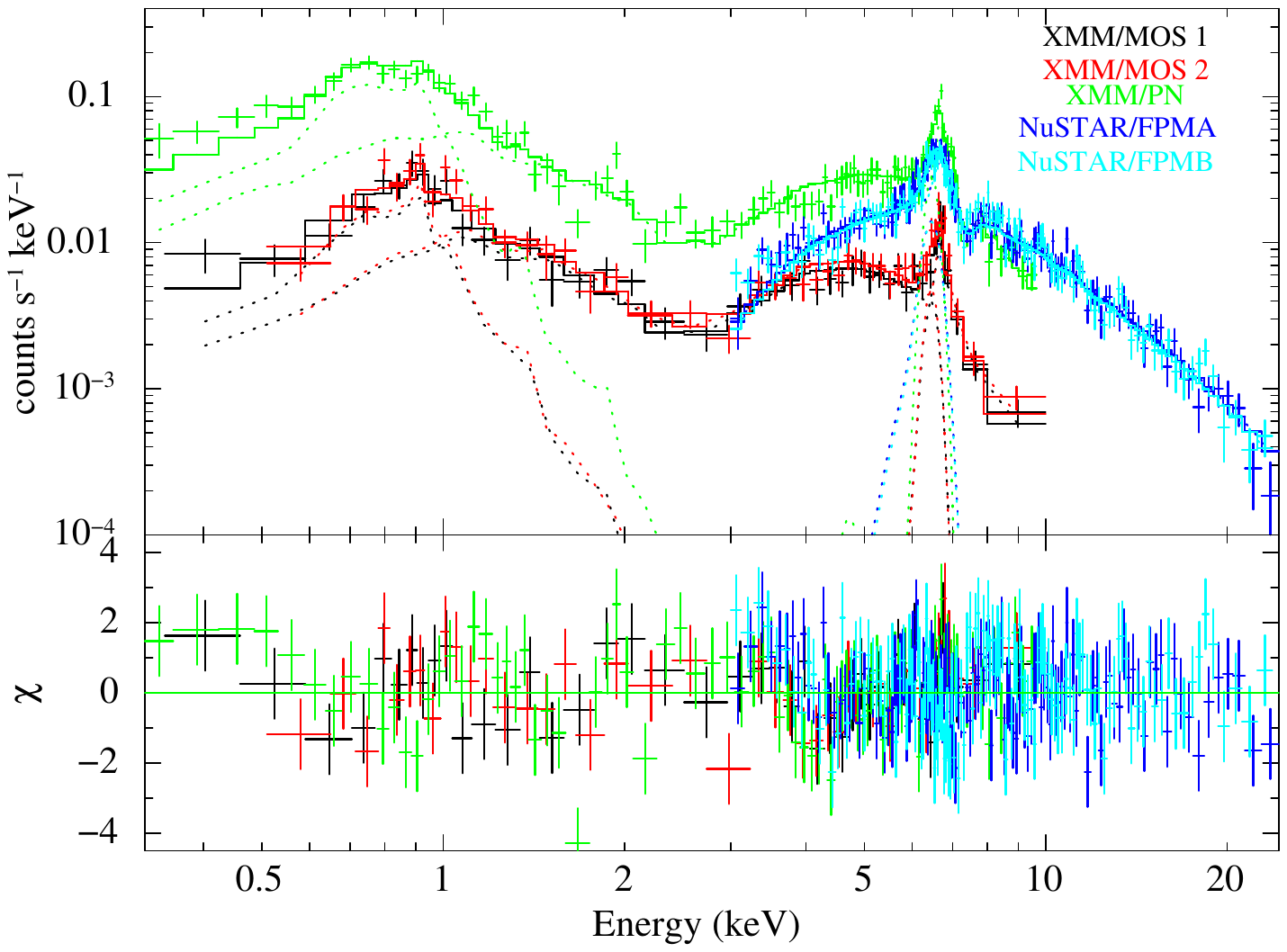}
    \includegraphics[scale=0.3]{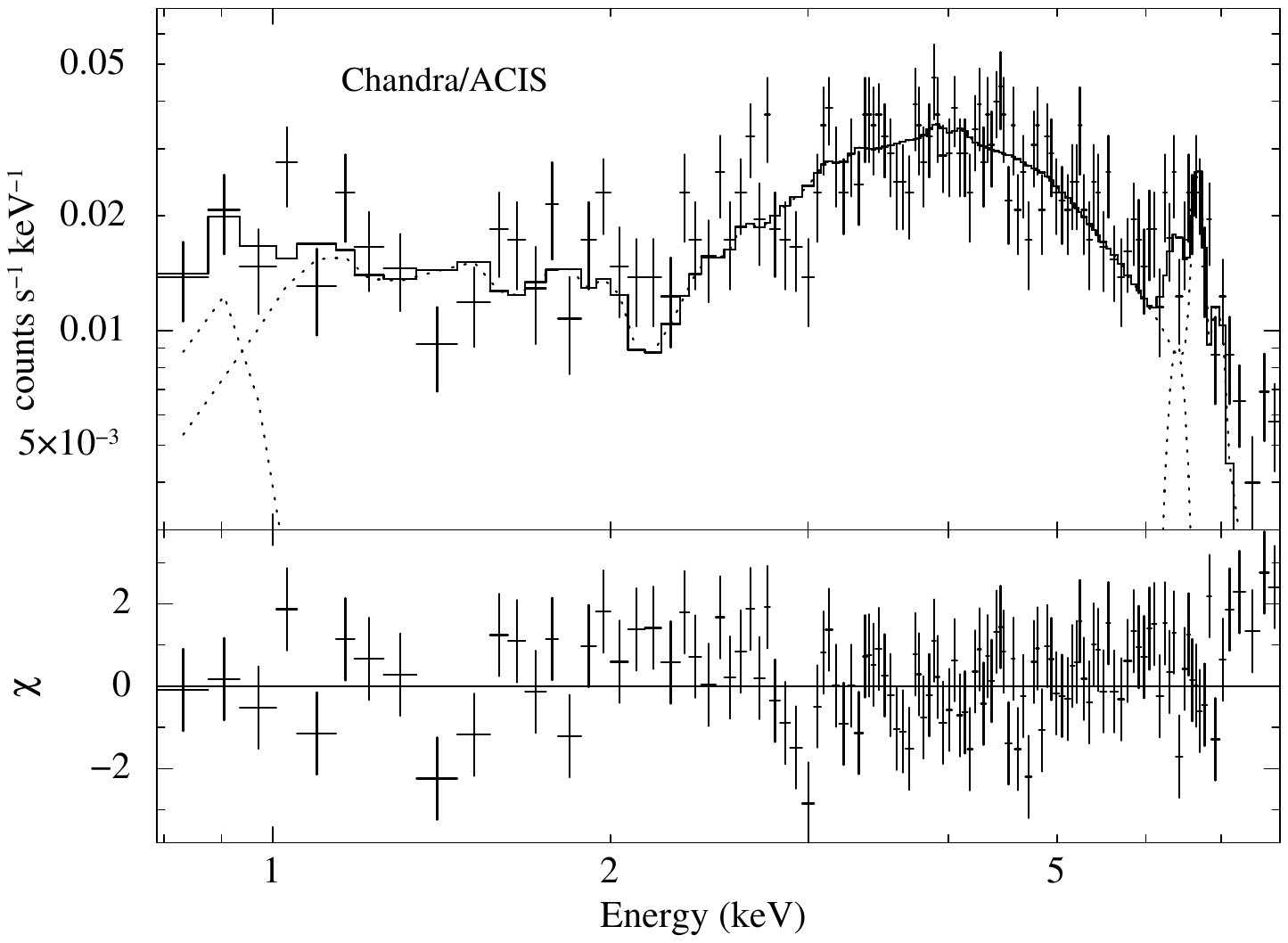}
    \caption{{\it Left panel}: Broadband X-ray spectral fits using XMM/EPIC and NuSTAR observations in 0.3--10 keV and 3--25 keV respectively, using the spectral model described in Section 2.4. {\it Right panel}: X-ray spectral fits to the Chandra/ACIS spectrum in 0.5--8.0 keV. The residuals to the spectral fits are plotted in the bottom panels. The spectral parameters of the best fit model are tabulated in Table 2.}
    \label{nustar_spec}
\end{figure}

\begin{table*}
\centering
\caption{Best fitting parameter values for the Chandra/ACIS and joint XMM/EPIC and NuSTAR spectra. The errors on the parameters are estimated using 90\% confidence limits}
\begin{tabular}{l c c}
\hline
Parameter & Chandra/ACIS & XMM/EPIC + NuSTAR  \\
& (0.3-8.0 keV) & (0.3-25 keV) \\
\hline
$C_{\rm{PN}}$ & - & 0.52$\pm$0.03  \\
$C_{\rm{MOS1}}$ & - & 0.34$\pm$0.02  \\
$C_{\rm{MOS2}}$ & - & 0.34$\pm$0.02  \\       
$C_{\rm{FPMA}}$ & - & 1.0 (fixed)  \\
$C_{\rm{FPMB}}$ & - & 1.02$\pm$0.04  \\
$N_{\rm{H_{1}}}$ (10$^{22}$cm$^{-2}$) & 0.2(fixed)  & 0.2 (fixed)  \\
$N_{\rm{H_{2}}}$ (10$^{22}$cm$^{-2}$) & 14$\pm$1 & 30$\pm$2  \\
CvrFrac & 0.965$\pm$0.005 & 0.958$\pm$0.005 \\
{\tt coolflow} kT$_{max}$ (keV) & 18(fixed) & 18$\pm$2  \\
{\tt coolflow} Normalization (M$_\odot$/yr) & (3.3$\pm$0.3)$\times 10^{-11}$ &  (4.3$\pm$0.8)$\times 10^{-11}$ \\
{\tt vapec} kT (keV) & 0.08$^{+0.13}_{-0.08}$ & 0.29$\pm$0.03  \\
{\tt vapec} Normalization & 0.09$^{+0.06}_{-0.08}$ & (1.3$\pm$0.3)$\times 10^{-4}$ \\
Abundance ({\tt coolflow}) Z$_{\odot}$ & 1.6$\pm$0.5 & 1.3$\pm$0.3 \\
Ne abundance ({\tt vapec}) Z$_{\odot}$ & 1.6(fixed) & 4$\pm$1 \\
Fe K$\alpha$ keV & 6.4(fixed) & 6.47$\pm$0.04 \\
$\sigma$ (Fe K$\alpha$) keV & 0.1 (fixed) & 0.14$\pm$0.05 \\
Norm (Fe K$\alpha$) photons/cm$^{2}$/s & (2.2$\pm$0.9) $\times 10^{-5}$ & (4.5$\pm$0.9) $\times 10^{-5}$ \\
Eqw (Fe K$\alpha$) keV & 0.18$\pm$0.09 & 0.30$\pm$0.04 \\
$\chi^{2}$ & 148.07 for 112 d.o.f & 506.48 for 434 d.o.f \\
Absorbed Flux 0.3-10.0 keV (10$^{-12}$ ergs/s/cm$^{2}$) & - & 4.4$\pm$0.1  \\
Absorbed Flux 0.5-8.0 keV (10$^{-12}$ ergs/s/cm$^{2}$) & 4.9$\pm$0.2 & - \\
Absorbed Flux 3-25 keV (10$^{-12}$ ergs/s/cm$^{2}$) & - & 6.8$\pm$0.1 \\
Unabsorbed Flux 0.3-10.0 keV (10$^{-12}$ ergs/s/cm$^{2}$) & - & 13.3$\pm$0.3 \\
Unabsorbed Flux 0.5-8.0 keV (10$^{-12}$ ergs/s/cm$^{2}$) & 18.9$\pm$0.7 & - \\
Unabsorbed Flux 3-25 keV (10$^{-12}$ ergs/s/cm$^{2}$) & - & 11.6$\pm$0.4 \\
\hline
\end{tabular}
\end{table*}

Broadband X-ray spectral fitting was carried out in the energy range 0.3--10 keV for XMM-EPIC MOS1, MOS2 and PN and 3--25 keV for NuSTAR FPMA and FPMB. We also fit separately the Chandra-ACIS spectra in the 0.5--8.0 keV energy range. In CVs, the emission from the X-ray emitting regions such as a post-shock plasma or a boundary layer can be modeled using a multi-temperature plasma model over a continuous temperature distribution, from the shock temperature to the WD photospheric temperature (T $\sim 10^{4}$ K) such as the cooling flow model ({\tt coolflow}) \citep{mukai2003, Islam2025}. Even the magnetic CVs that \cite{mukai2003} classified as having a photo-ionized spectrum are now understood to have a cooling-flow like primary emission absorbed by a complex absorber (see, {\it e.g.}, \citealt{islam2021}). The old nova, V603~Aql (which, by definition, is a nova-like system in terms its accretion-powered properties) is one of the four  systems originally recognized as having a cooling-flow type X-ray spectrum \citep{mukai2003}. For this reason, we apply the cooling flow model to UU~Aqr, even though this model may not precisely reproduce the emission measure distribution of nova-like systems \citep{baskill2005}. The cooling flow model is modified by a fully covering photo-electric absorption model {\tt Tbabs}, a partial covering absorption model {\tt Tbpcf} and a Gaussian line for the strong 6.4 Fe K$\alpha$ fluorescence line. The redshift of the cooling flow model is fixed to the cosmological redshift of 6$\times 10^{-8}$ derived from converting the Gaia distance of the source assuming the $H_{\rm 0}$ = 70 km/s/Mpc. The absorption column density $N_{\rm H}$ for the fully covering absorption model is fixed to the interstellar $N_{\rm H}$ of 2 $\times 10^{21}$ cm$^{-2}$, the X-ray absorption column density value taken from 3D extinction map of \cite{doroshenko2024} based on the optical-IR reddening of stars with Gaia distances and the canonical relation between the extinction co-efficients and interstellar $N_{\rm H}$ \footnote{http://astro.uni-tuebingen.de/nh3d/nhtool}.
We initially kept $kT_{\rm{max}}$ of the {\tt coolflow} model free while fitting the Chandra spectra. However the spectral fit and the error estimation could not constrain it to a reasonable value, since the upper energy range of the Chandra/ACIS spectra is 8 keV. Hence we fix it to the value estimated from the joint XMM and NuSTAR spectral fits.  To account for the soft excess seen in the XMM and Chandra spectra at lower energies $\leq$ 1 keV, we model it with an additional collisional ionized plasma model {\tt vapec}, which is modified by only the fully covering absorption model {\tt Tbabs}. We set the abundance table of the total and partial covering absorption model {\tt Tbabs} and {\tt Tbpcf}, cooling flow model {\tt coolflow} and {\tt vapec} model to {\tt aspl} abundance table \citep{asplund2009}. The abundance of all elements except Ne in the {\tt vapec} model is tied to its abundance in the {\tt coolflow} model. The Ne abundance is kept free for the joint XMM and NuSTAR spectral fits but is kept fixed to 1.6 (which is the abundance estimated from the {\tt coolflow} model) for the Chandra spectral fit. The XMM data strongly suggest a super-solar abundance of Ne, which was also seen in V603~Aql \citep{mukai2005}. A constant was added to the XMM MOS1, MOS2, PN and NuSTAR FPMA and FPMB spectral fits to account for the cross-normalization difference between the different instruments. All spectral fits were carried out using {\tt XSPEC v12.15.1}  \citep{arnaud1996}. The full spectral model in {\tt XSPEC} is {\tt constant*(Tbabs*(Tbpcf*(coolflow+gaus))+ vapec)}. 
Table~2 provides the spectral parameters of the best fit to the Chandra spectra and the joint fits to the simultaneous XMM and NuSTAR spectra. The absorbed X-ray luminosities in the soft (0.3-10 keV) and hard (3-25 keV) energy bands are 3.4$\times 10^{31}$ ergs/s and 6.1$\times 10^{31}$ ergs/s respectively, whereas the unabsorbed X-ray luminosities are an order of magnitude higher. The unabsorbed X-ray luminosity in 0.3-25 keV energy band of only the hard component {\tt coolflow} is 1.2$\times 10^{32}$ ergs/s, whereas the unabsorbed X-ray luminosity in 0.3-25 keV energy band of only the soft {\tt vapec} component is 1.3$\times 10^{30}$ ergs/s.

\section{Results and Discussion}

\begin{figure}
    \centering
    \includegraphics[scale=0.5]{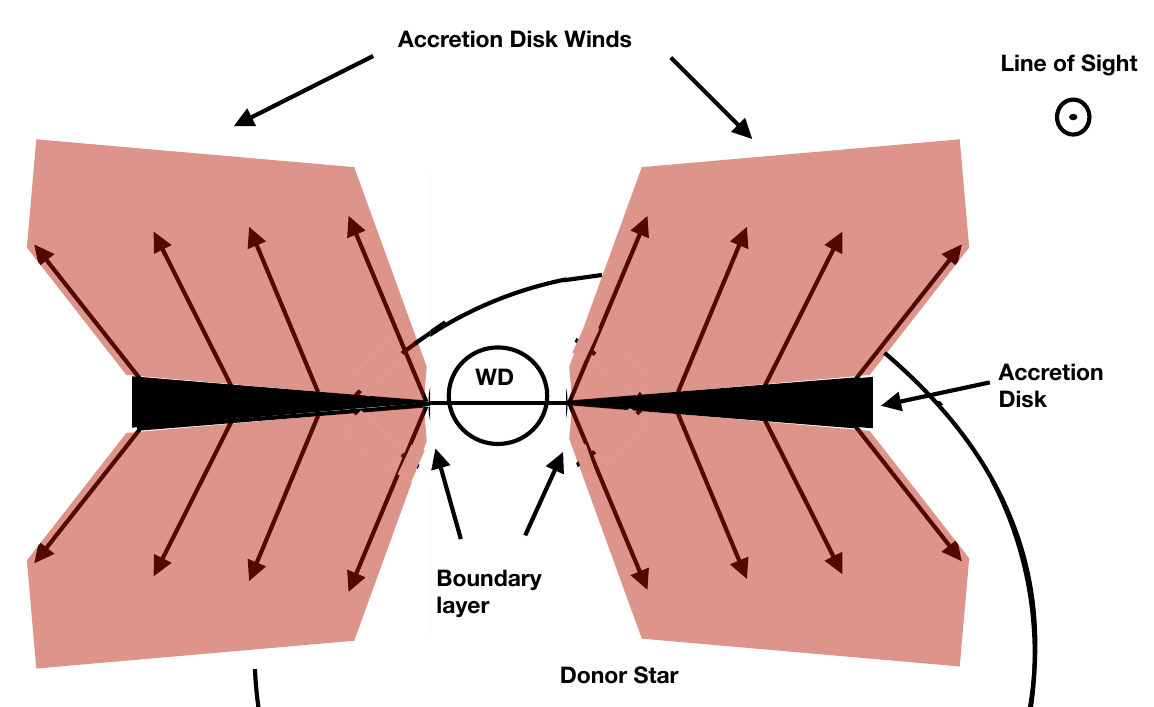}
    \caption{Schematic diagram of UU~Aqr. The line of sight of the observer is into the page and denoted by a vector notation of a circle with a dot. The eclipsed hard X-ray emission is likely produced in the boundary layer between the accretion disk and the WD, situated on the orbital plane. In contrast, the soft, un-eclipsed X-ray emission likely originates from accretion disk winds situated above the orbital plane. The parameters for the accretion disk winds are taken from \cite{knigge1997} for another eclipsing nova-like variable UX~UMa. }
    \label{diagram}
\end{figure}

From Figure \ref{orb_phase}, we confirm the presence of a total X-ray eclipse in the hard energy band 3--10 keV XMM-EPIC and in 3--25 keV NuSTAR orbital intensity profiles. No X-ray eclipse is detected in the soft energy band (0.3–2 keV) of the XMM-EPIC lightcurves, while the Chandra-ACIS lightcurve in the 0.3–8 keV energy band suggests a tentative partial X-ray eclipse. The energy dependence of the eclipse behavior indicates that UU~Aqr is the second nova-like system (after UX~UMa) with two distinct X-ray components that can be clearly distinguished using both by their lightcurves and X-ray spectra (Figure \ref{nustar_spec}). From the orbital intensity profiles and X-ray spectra, we notice the hard component (dominant above 3 keV) is absorbed and eclipsed, while the soft component (dominant below 2 keV) is unabsorbed and un-eclipsed. Moreover, the left panel of Figure \ref{pow_spec} confirms that these two X-ray components of UU~Aqr also differ in terms of their aperiodic variability characteristics: the hard component is found to be variable in the frequency range 10$^{-3}$--10$^{-1}$ Hz (10--1000 s) while the power spectrum of the soft component is consistent with Poisson noise (left panel of Figure \ref{pow_spec}).

\subsection{The Hard X-ray Component}

As shown in Figure \ref{eclipse_fit_asym}, the orbital intensity profiles constructed from XMM-EPIC 3--10 keV and NuSTAR 3--25 keV lightcurves have flat-bottomed eclipse profiles and are consistent with a complete eclipse of the hard X-ray component. In a total X-ray eclipse, the mid-ingress to mid-egress interval reflects the projected size of the secondary where it crosses in front of the WD, while the ingress and egress durations reflect the size of the X-ray emitting region. Unfortunately, the measured ingress and egress durations either have large errors ({\it e.g}, 0.005$\pm$0.005 in the case of egress duration as measured with XMM-EPIC 3--10 keV data), or are inconsistent with each other (0.010$\pm$0.003 and 0.004$\pm$0.001 for the ingress duration for the NuSTAR and XMM-EPIC 3-10 keV eclipse). These eclipse profiles are influenced by the source variability unrelated to the eclipse, as seen by the difference between the pre-eclipse and post-eclipse count-rates measured in the XMM-EPIC 3--10 keV data.
\par
Nevertheless, we can constrain the size of the X-ray emission region using the observed ingress and egress durations. For this, we assume the mass of the WD of 0.67~M$_{\odot}$, a donor star mass of 0.2~M$_{\odot}$, and an orbital separation of 3.8 $\times 10^{10}$ cm, as estimated using optical observations by \cite{baptista1994}. The combined motion of the WD and the donor is 2$\pi \times 3.8 \times 10^{10}$ = 2.4 $\times 10^{11}$ cm over one orbital cycle. The orbital intensity profiles constructed with NuSTAR lightcurves are folded over five orbital cycles, and therefore averages over the short term variability compared to the single orbital cycle covered by the XMM-EPIC lightcurves. Thus, the NuSTAR ingress duration of 0.01 corresponds to a size of 2.4 $\times 10^{9}$ cm, about 50\% larger than the diameter of a 0.67 M$_{\odot}$ WD of 1.6 $\times 10^{9}$ cm. If we use the XMM-EPIC ingress duration of 0.004 instead, then the estimated X-ray emission region size is smaller than that the WD. An emission region smaller than the primary is unphysical; within the uncertainties the emission region appears to be slightly larger than the WD, similarly to what has been inferred in eclipsing DNe OY~Car \citep{ramsay2001,wheatley2003b}, HT~Cas \citep{nucita2009}, and Z~Cha \citep{nucita2011} in quiescence. This suggests that the hard X-ray component likely originates in the boundary layer between the Keplerian accretion disk and a much more slowly rotating WD. 
\par
The XMM-EPIC and NuSTAR spectrum of the hard X-ray component is described using a cooling flow model modified by a fully covering and a partial covering absorption, with a Gaussian model for the Fe K$\alpha$ fluorescence line. The fluorescence line is considered to originate from reflection on the WD surface, which is expected for a compact emission region in close proximity. The column density of the partial covering absorption component is of order $10^{23}$ cm$^{-2}$.  We interpret this as due to the fact that our line of sight to the emission region passes through the outer layer (``atmosphere'') of the the accretion disk, which is consistent with a boundary layer origin of the hard X-ray component.
\par
The best-fit maximum temperature ($kT_{\rm {max}}$) of the cooling flow component is 18$\pm$2 keV. If the boundary layer plasma in UU~Aqr has the highest possible temperature, converting all the kinetic energy of a Keplerian flow just above the WD surface ({\it i.e}, half the total potential energy) to thermal energy in a strong shock, the observed $kT_{\rm {max}}$ of 18 keV requires a WD mass of 0.8 M$_{\odot}$ \citep{byckling2010}. We note that this mass estimate of 0.8 M$_{\odot}$ is a lower limit, since $kT_{\rm {max}}$ of non-magnetic CVs appear lower than the strongest shock from Keplerian flow \citep{mukai2022}. The observed $kT_{\rm {max}}$ in quiescent DNe suggests actual $kT_{\rm {max}}$ is somewhat lower than this theoretical maximum, with a value near 12--14 keV for a 0.8 M$_{\odot}$ WD \citep{yu2018}. Furthermore, if the boundary layer of nova-like systems is similar to that of DNe in outburst, the temperature is expected to be lower still for the same mass \citep{wheatley2003a}. Therefore, there is a discrepancy between the $kT_{\rm {max}}$ and the estimated WD mass of 0.67$\pm$0.14 M$_{\odot}$ \citep{baptista1994}.
\par
One possibility could be that the estimated WD mass from the optical observations is incorrect. A potential cause for this discrepancy could be due to the fact that \cite{baptista1994} derived the WD mass by associating the compact light source, seen in the optical eclipse lightcurves, with the WD photosphere. As mentioned in \cite{baptista1994} (see also \citealt{wood1986,baptista1995}), the compact light source may turn out to be a combination of the WD photosphere and an optically bright boundary layer. We are not aware of any systematic effects that would allow a $<$0.8 M$_{\odot}$
non-magnetic WD to achieve a high $kT_{\rm {max}}$ of 18 keV. Given the higher WD mass implied by the measured $kT_{\rm {max}}$ of the hard X-ray component, a reassessment of the system parameters of UU~Aqr by new optical/UV observations could be necessary.
\par
Using the duration of the X-ray eclipse, we can also provide estimates on the radius of the secondary star. From the eclipse fits to the NuSTAR profile in Table 1, we estimate an eclipse half-angle $\theta_{E}$ of 8.8$^{\circ}$. The relation between the eclipse half-angle $\theta_{E}$ and radius of the donor star R$_{0}$ is given below, where {\it a} is the orbital separation.

\begin{equation}
\frac{R_0}{a} = \sqrt{\cos^2 i + \sin^2 i \, \sin^2 \theta_{\rm E}}
\label{eq:geometry}
\end{equation}

The inferred inclination {\it i} from optical observations is 78$^{\circ}$ by \cite{baptista1994}. However this estimate is highly dependent on the modeling the eclipse lightcurves and have large uncertainties. There we assume {\it i} to have a range from 75 to 80$^{\circ}$. We assume an empirical relation between masses and radius of the low mass Zero-Age Main Sequence stars (ZAMS) in CVs from \cite{patterson1984} given below

\begin{equation}
\frac{R_0}{R_\odot} = \left( \frac{M}{M_\odot} \right)^{0.88}
\label{eq:mass-radius}
\end{equation}

We estimate the mass of the donor star to be 0.28 to 0.29 M$_{\odot}$ and radius to be 0.33 to 0.34 R$_{\odot}$. The mass ratio {\it q} estimated from the optical measurements of eclipses by \cite{baptista1994} is 0.3. The mass-radius ($M_{wd}$--$R_{wd}$) relation for WD from \cite{hamada1961} is given below, where $M_{ch}$ is the Chandrasekhar mass limit for WD.

\begin{equation}
\frac{R_{\rm wd}}{R_\odot}
= 0.0112
\left[
\left( \frac{M_{\rm wd}}{M_{\rm Ch}} \right)^{-2/3}
-
\left( \frac{M_{\rm wd}}{M_{\rm Ch}} \right)^{2/3}
\right]^{1/2}
\label{eq:wd-mass-radius}
\end{equation}

We estimate the mass of the WD to be in the range of 0.94 to 0.97 M$_{\odot}$ and radius of the WD to be 5523 to 5797 km. This estimate of the mass of the WD is based on the assumption that the secondary of UU~Aqr obeys the empirical mass-radius relationship of \cite{patterson1984} and is consistent
with the mass estimate from the observed $kT_{\rm {max}}$ of the cooling flow model. This new mass estimate of the WD is consistent, within the associated uncertainties, with the average masses of WD in non-magnetic CVs estimated by \cite{Yu2022}. 

\subsection{The Soft X-ray Component}

The soft X-ray component lacks an X-ray eclipse, as observed in the XMM-EPIC orbital intensity profile in 0.3--2.0 keV (second panel of Figure \ref{orb_phase}). This is in stark contrast not only to the eclipses in the hard X-ray component but also to the eclipses in the optical/UV component. To explore the implications of a lack of soft X-ray eclipse, we first briefly discuss the eclipse profile of the XMM-OM UVW1 lightcurve.
\par
The average of the derived ingress and egress duration estimated from fitting the XMM-OM UVW1 lightcurves is about 0.09 in orbital phase, and corresponds to a UV emitting region of size $2.4 \times 10^{10}$~cm using the same procedure as used for the hard X-ray component, or about 0.45 times the orbital separation. This is consistent with an accretion disk origin for the near UV band (UVW1 filter), and suggests that the UV-emitting part of the disk must occupy a large fraction of the primary Roche lobe. A distinctly V-shaped eclipse is clearly detected in the XMM-OM data, in contrast to a lack of eclipse in the soft X-ray component.
\par
However, there is an un-eclipsed component of the optical emission from UU~Aqr, including emission lines \citep{baptista2000}, which is interpreted as lines originating mostly in an accretion disk wind. While a region extended solely within the orbital plane would need to be several times larger than the binary separation to avoid even a partial eclipse (as inferred from the OM data), a vertically extended region can remain un-eclipsed. It is reasonable to assume that the soft X-ray emission and the un-eclipsed optical emission lines originate from the same general region, most likely a vertically extended, radiatively driven accretion-disk wind, with the X-rays produced in shocks associated with the well-known instabilities of line-driven winds. \citep{Lucy1970, Lucy1980, Owocki1988, Feldmeier1995}.
\par
We compare the timing characteristics of the hard and soft X-ray bands, to further probe the nature of the soft X-ray emission. The left panel of Figure \ref{pow_spec} displays the power spectra of UU~Aqr constructed from the soft and hard X-ray energy bands of XMM-EPIC and NuSTAR. We see that the hard X-ray band is variable at frequencies below about 10$^{-2}$ Hz, while any variability at higher frequencies is below the Poissonian level of these power spectra.  In contrast, we see no clear variability in the soft band at any frequencies. In UX~UMa, the other nova-like system with distinct two-component X-ray emission, the situation is a little different. Both the hard and the soft X-ray bands show a power excess (modeled at power-law) at low frequencies up to $\sim 10^{-2}$ Hz. There is an additional component of variability in the hard X-ray band, centered around $2 \times 10^{-3}$ Hz which we modeled with a Gaussian, which is absent in the soft X-ray band. The  variability of the hard component is not unexpected and similar variability has been studied in several other nova-like systems (see, {\it e.g.}, \citealt{dobrotka2019}). In this study, we concentrate only on the different timing characteristics of the hard and soft components.
\par
We investigate the possibility of cross-contamination of the hard component to the soft band X-ray flux. The X-ray spectra at E $<$ 2 keV is modeled by a collisional ionized plasma model and has majority of its contribution below 2 keV, the contribution of the soft component to the hard band (E $>$ 3 keV) is likely to be negligible. However, the hard component could potentially contribute to the soft X-ray flux (E $<$ 2 keV) through the intrinsic partial covering absorber.  In the case of UX~UMa, we estimate that the hard component contributes about 10\% of the absorbed X-ray flux in the soft band, based on the best-fit spectral parameters. In the case of UU~Aqr, the intrinsic absorption is higher than in UX~UMa (Figure \ref{nustar_spec}), and we estimate that the contribution of the hard component to the soft band X-ray flux is even lower (of order 1\%).
\par
For UX~UMa, \cite{pratt2004} proposed the scattering of the hard X-ray component in an extended region as the possible origin of the soft component.  However, the fact that there is significant (greater than the Poisson level) power in the hard component near 10$^{-3}$ Hz that is not present in the soft component is a significant concern for this interpretation. Scattering itself should be instantaneous, thereby preserving the variability characteristics of the direct emission in the scattered photons. For the $\sim$1000 s variability of the hard component to be smeared out by the light-travel time within the scattering cloud, the size of that cloud must be of order 1000 light seconds or larger, far away from the binary itself which has a size of a few light seconds. While the accretion disk wind may well extend that far, it would be puzzling if the scattering only happened well away from the binary and not in the immediate vicinity of the binary.
\par
Geometrically, we suggest a possible association of soft X-ray emission and the accretion disk wind. Spectrally, the soft component is similar to the beta-type component in symbiotic stars \citep{luna2013}. In this case the beta component is usually interpreted as due to colliding winds -- an accretion disk wind colliding with the massive, slow wind from the red giant mass donor. Since the secondary star in UU~Aqr is a low mass main sequence star, the scenario involving the interaction between a red giant wind and the accretion disk wind is not applicable in this system. We therefore consider if an internal shock within the accretion disk wind can explain the observed soft component. Previously, an internal shock model was proposed for the beta-type X-ray component of the symbiotic star MWC 560 because optical spectroscopy indicates a collimated outflow in that system \citep{lucy2020}.
\par
In the case of early-type stars, both binary systems ({\it e.g}, $\eta$ Car and WR 140) and single stars are known to be X-ray sources, with binaries typically exhibiting higher temperatures and X-ray luminosities. In single stars, internal shocks in  single star winds convert a fraction of the kinetic power (W$_{wind}$) to X-rays.  We note that both early type stars and accretion disks in CVs and symbiotic stars have line-driven winds that require non-uniform outflow velocities to maintain resonant scattering in the lines without producing a self-shadowing effect, thus allowing radiation pressure to drive outflows at much less than the Eddington luminosities. This makes them subject to the same line-deshadowing instability \citep{feldmeier2003} that is responsible for the X-rays originating in the winds of early-type stars.
\par
Early type stars where X-rays are dominated by internal shocks are empirically found to have $L_{x} / L_{bol} \sim 10^{-7}$ \citep{Harnden1979,Long1980,Pallavicini1981,Naze2009}. \cite{Owocki2013} found that this relation can be understood in terms of thin-shell mixing in radiatively cooled shocks. In this model, the X-ray luminosity is expected to scale as $L_{x} \propto (\dot{M} / v_\infty)^{1-m}$, where $m=0.2-0.4$, depending on assumptions regarding the scaling of wind density with bolometric luminosity. At lower wind densities, radiative cooling becomes inefficient, and adiabatic cooling dominates, and asymptotically $L_{x} \propto (\dot{M} / v_\infty)^2$.
\par
To interpret the X-ray luminosities of UX~UMa and UU~Aqr in this paradigm, we should ideally consider the following factors: the radiation field is different (a single temperature photosphere for the stars and a multi-temperature disk, WD photosphere plus the disk-WD boundary layer for the CVs); the geometry is different (spherical vs. bi-conical); perhaps most importantly, the effective gravity is different (note that the accretion disk wind is launched from a rapidly rotating accretion disk whose Keplerian motion effectively cancels the gravity of the WD). In the absence of further development of the theory of \cite{Owocki2013} for accretion disk winds, we compare the observed properties of UX~UMa and UU~Aqr to those of O stars in the context of the scaling law for radiatively cooled shocks. The mass loss rate and wind terminal velocity of the O supergiant $\zeta$ Pup are 3.5$\times$10$^{-6}$ M$_\odot$\,yr$^{-1}$ \citep{cohen2010} and 2250 km\,s$^{-1}$ \citep{Puls2006}, while its X-ray luminosity is $1.85 \times 10^{32}$ ergs\, s$^{-1}$ \citep{Naze2009}, assuming a distance of 335 pc \citep{MaizApellaniz2008}.
\par
Unfortunately, there appears to be no estimate in the literature of the mass loss rate for the wind in UU~Aqr. Therefore we concentrate on the case of UX~UMa.  Using the soft component un-eclipsed X-ray flux of 2.9$\times$10$^{-13}$ ergs\,cm$^{-2}$s$^{-1}$ extrapolated to 0.2--10 keV (which does not include the hard component above 2 keV) \citep{pratt2004} and the Gaia distance of 292 pc, the L$_{x,uneclipsed}$ is about 3$\times 10^{30}$ ergs\,s$^{-1}$ . \cite{noebauer2010} used a wind mass loss rate of 10$^{-9}$ M$_\odot$\,yr$^{-1}$, or $\sim 6.3 \times 10^{16}$ g\,s$^{-1}$. We adopt a wind terminal velocity of 3000 km\,s$^{-1}$ as a reasonable estimate, perhaps even a lower limit, based on \cite{mason1995}, who equated the maximum blueshift observed in emission lines of $\sim$ 2700 km\,s$^{-1}$ with the wind terminal velocity. While this inference is correct for a spherical wind, accretion disk winds are thought to be bi-conical (see, {\it e.g.}, Figure 5 of \citealt{noebauer2010} for two versions of a schematic).  Given this, the value of 2700 km\,s$^{-1}$ is an underestimate by an unknown projection factor. 
Scaling the X-ray luminosity of $\zeta$
 Pup by $(\dot{M} / v\infty)^{0.6}$ (assuming $m=0.4$) leads to a predicted luminosity of $1.2\times10^{30}$ ergs\,s$^{-1}$. If this estimate is correct, the accretion disk wind of UX~UMa needs to be a factor of a few more efficient at generating X-rays than that of $\zeta$ Pup for the internal shock model of soft X-rays to be viable. The wind parameters in UU~Aqr are likely to be similar to those of UX~UMa, while the soft un-eclipsed X-ray luminosity is 1.3 $\times 10^{30}$ ergs\,s$^{-1}$, comparable to the estimate scaled from $\zeta$ Pup.
\par
However, the wind mass loss rate of CVs is notoriously uncertain. Specifically, \cite{tampo2024} modeled the optical emission lines of V455 And during its superoutburst as originating in the disk wind, and estimated a wind mass-loss rate 1--2 orders of magnitude higher than the previous CV wind models. Further experimentation with their model led them to conclude that the normalized spectra and P-Cygni profiles in UV wavelengths are much less sensitive to the mass-loss rate than the continuum, suggesting that previous wind models have been likely biased towards lower wind mass-loss rates and that winds similar to that in V455 And may be present in some nova-like variables, and noted characteristics of the Balmer lines of some nova-like systems that are similar to those seen in V455 And superoutburst.  If the revised wind mass loss rate estimate of \cite{tampo2024} is correct, then the soft X-ray component of UX~UMa, and by analogy, UU~Aqr, may well be due to internal shocks within the accretion disk wind. 
\par
On the other hand, if the accretion disk winds of UX~UMa and UU~Aqr are in the adiabatic regime, or the intermediate regime between radiative and adiabatic, our previously predicted X-ray luminosities have been overestimated. Higher mass-loss rates would mitigate the discrepancy both by directly increasing the X-ray luminosity prediction and by increasing the relative importance of radiative cooling.
\par
Figure \ref{diagram} depicts the schematic diagram of UU~Aqr. The eclipsed hard X-ray emission is thought to originate from the the boundary layer between the accretion disk and the WD whereas the soft un-eclipsed X-ray emission is considered to be from the vertically extended accretion disk winds. For the schematic diagram, the WD and the donor star are scaled appropriately from their system parameters and the parameters for the accretion disk winds are taken from \cite{knigge1997}, which are estimated for another eclipsing nova-like variable UX~UMa.

\subsection{Wider Implications}

We have discovered two distinct X-ray components in UU~Aqr, the second nova-like system (after UX~UMa) to display such characteristic. We suggest, however, that this reflects the fortuitous viewing geometry (UX~UMa has an estimated inclination angle of 75$^\circ$ and UU Aqr of about 78$^\circ$), and that several nova-like systems likely have a compact hard X-ray source in addition to the extended soft X-ray source.  Since the former is more luminous, it is difficult to notice the latter as a distinct component, if it was not for the intrinsic absorber that attenuates the hard component below 2--3 keV.  This is possible when our line of sight to the boundary layer passes through the accretion disk atmosphere. For even higher inclination angles, our line of sight to the boundary layer passes through the body of the accretion disk, completely hiding the hard X-ray component from our view: this could explain the observation of OY Car in superoutburst \citep{naylor1988,pra99b}.
\par
At lower inclination angles, the presence of two distinct components may not be readily apparent, which could lead to modeling the observed X-ray spectra with a multi-temperature model in which the contribution from lower-temperature plasma is enhanced relative to that expected from a pure cooling-flow scenario. This is indeed the case, at least qualitatively, for DNe in outburst and for nova-like systems \citep{baskill2005}.  It remains to be seen if this explanation holds up quantitatively.
\par
It is also worth re-examining if the $\beta$-type component seen in some symbiotic stars is indeed due to colliding winds, as is widely thought. A head-on collision between the accretion disk wind and the massive, slow outflow from the late type giant is highly efficient at converting the wind kinetic energy to X-rays. This efficiency may in fact be too high, at least with respect to the resulting X-ray temperature: a strong shock produced by a wind with a velocity of 3000 km\,s$^{-1}$ would be expected to yield a high plasma temperature (kT$\sim$10 keV), whereas the defining characteristic of the $\beta$-type component is its relatively modest temperature (kT$\sim$ 1 keV), comparable to that observed from internal shocks in the winds of single early-type stars. It is likely that the bi-conical geometry of the accretion disk wind, combined with the equatorial concentration of M-giant outflow in symbiotic binaries (see, {\it e.g.}, \citealt{booth2016}), prevents head-on collisions from dominating the observed X-ray emission.  If this is the case, then the $\beta$-type component seen in symbiotic stars may also be due to internal shocks within the accretion disk wind.

\section*{Acknowledgement}
We thank the anonymous referee for the constructive
comments that helped improve the manuscript. We thank Felix Fuerst for useful discussions. The scientific results reported here are based on observations made by the NuSTAR X-ray observatory, and we thank the NuSTAR Operations, Software, and Calibration teams for scheduling and the execution of these observations.
This research has made use of data obtained from the Chandra Data Archive provided by the Chandra X-ray Center (CXC). This paper employs a list of Chandra datasets, obtained by the Chandra X-ray Observatory, contained in the Chandra Data Collection ~\dataset[DOI: 10.25574/cdc.542]{https://doi.org/10.25574/cdc.542}. Based on observations obtained with XMM-Newton, an ESA science mission with instruments and contributions directly funded by ESA Member States and NASA. Support for this work was provided by NASA through grant No. 80NSSC24K1542. ML acknowledges support from NASA's Astrophysics Division. GWP acknowledges long-term support from CNES, the French space agency.

\software{XSPEC (v12.15.0; \citealt{arnaud1996}); Scipy (v1.15.2; \citealt{2020SciPy-NMeth})
}

\bibliographystyle{aasjournalv7}
\bibliography{bibtex}
\end{document}